\documentclass[%
aps,
jmp,%
amsmath,amssymb,
reprint,%
]{revtex4-2}
\usepackage{graphicx}
\usepackage{subfigure}
\usepackage{dcolumn}
\usepackage{bm}
\usepackage[mathlines]{lineno}
\usepackage{graphicx}

\usepackage{graphicx}
\usepackage{subfigure}
\usepackage{dcolumn}
\usepackage{bm}
\usepackage[mathlines]{lineno}
\graphicspath{{figure/}}
\usepackage{array}
\usepackage{braket}
\usepackage{diagbox}
\usepackage{amsmath}
\usepackage{appendix}
\usepackage{multirow}
\usepackage{bm,color,bbm}

\newcommand{\blk}{\color{black}}

\usepackage{float}
\newcolumntype{.}{D{.}{.}{-1}}
\usepackage{amsopn}
\usepackage[colorlinks,linkcolor=blue, anchorcolor=blue, urlcolor=blue, citecolor=blue]{hyperref}
\usepackage{epstopdf}

\begin{document}

	\title {Scalable twin-field quantum key distribution network enabled by adaptable architecture}
	\author{Chunfeng Huang$^{1}$, Rui Guan$^{1}$, Xin Liu$^{1}$, Wenjie He$^{1}$, Shizhuo Li$^{1}$, Hao Liang$^{1}$, Ziyang Luo$^{1}$, Zhenrong Zhang$^{2}$, Wei Li$^{3,4}$, and Kejin Wei$^{1}$}
	\address{
		$^1$Guangxi Key Laboratory for Relativistic Astrophysics, School of Physical Science and Technology, Guangxi University, Nanning 530004, China\\		
		$^2$Guangxi Key Laboratory of Multimedia Communications and Network Technology, School of Computer Electronics and Information, Guangxi University, Nanning 530004, China\\
        $^3$School of Artificial Intelligence Science and Technology, University of Shanghai for Science and Technology, Shanghai 200093, China\\
        $^4$Institute of Photonic Chips, University of Shanghai for Science and Technology, Shanghai 200093, China\\
	}

		\begin{abstract}

Quantum key distribution (QKD) is a key application in quantum communication, enabling secure key exchange between parties using quantum states. Twin-field (TF) QKD offers a promising solution that surpasses the repeaterless limits, and its measurement-device-independent nature makes it suitable for star-type network architectures. In this work, we propose a scalable TF-QKD network with adaptable architecture, where users prepare quantum signals and send them to network nodes. These nodes use an optical switch to route the signals to multi-user measurement units, enabling secure key distribution among arbitrary users and adapting to complex connection demands of the network. A proof-of-principle demonstration with three users successfully achieved secure key sharing over simulated link losses of up to $30$ dB, with an average rate of $19.57$ bit/s. Additionally, simulations show that the proposed architecture can achieve a total secure key rate of $4.84 \times 10^{4}$ bit/s at $100$ km in a symmetric $32$-user network. This approach represents a significant advancement in the topology of untrusted-node QKD networks and holds promise for practical, large-scale applications in secure communication.
		
		
			
    	\end{abstract}
	    \maketitle
	    
    	\section{Introduction}   	
        Quantum key distribution (QKD) is one of the most mature applications of quantum information processing~\cite{2018Wehner}. It uses quantum states to securely distribute symmetric keys between communication parties. Since the first QKD protocol was proposed~\cite{1984BEN}, significant efforts have steadily advanced QKD technology~\cite{2016Yin-404,2018Boaron-421,2023grunenfelder-fast,2023li,2023Li-Free-fiber,2025Li-Microsatellite,2025Zhang-MPbreak}. Recently, a record-breaking achievement was made, with secure key rates exceeding tens of megahertz~\cite{2023grunenfelder-fast,2023li}. Trusted-node QKD networks~\cite{2009Peev,2011Sasaki,2014Wang,2019Dynes,2021chen-implementation,2021chen-4600}, including integrated space-ground quantum communication networks~\cite{2021chen-4600}, have been rapidly advancing. However, challenges persist, particularly in long-distance transmission and network expansion, both of which are crucial for the wider deployment of QKD.
   	    	
    	\begin{figure*}[htp]
    		\centering	
    		\includegraphics[width=1\linewidth]{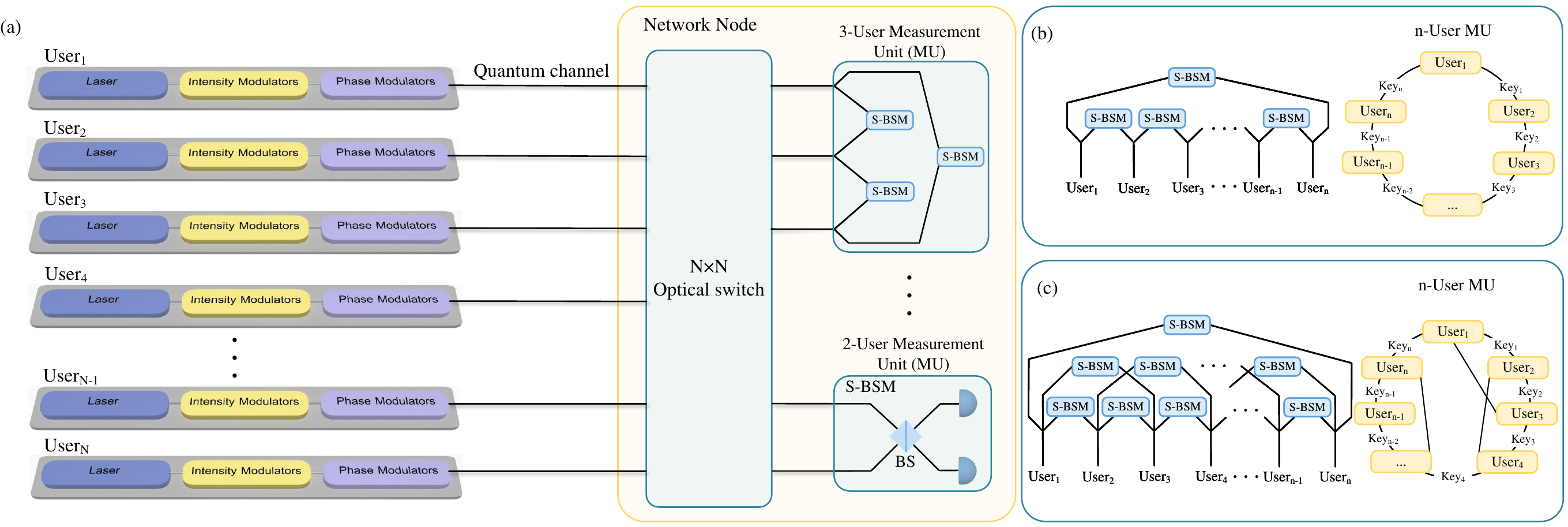}			
    		\caption{(a) Schematic of the scalable TF-QKD network architecture for $N$ users. The physical layer adopts a star topology consisting of end users, quantum channels, and untrusted network node. Each user requires only a transmitter. The network node utilizes an optical switch to manage user connections, directing quantum signals from users to designated measurement units (MUs). (b) Schematic of $n$-user MU with two extended ports per user output. (c) Schematic of $n$-user MU with three extended ports per user output. 
    		}
    		\label{fig_network_architecture}
    	\end{figure*}

        In previous protocols and systems~\cite{2014Lo-review,2016Diamanti}, the key generation rate decreases exponentially as the transmission distance increases. Without quantum repeaters, QKD is unable to overcome the fundamental rate-distance limits~\cite{2014Takeoka,2017Pirandola}. Fortunately, twin-field (TF) QKD~\cite{2018Lucamarini} provides a practical long-distance solution that exceeds the repeaterless limit. Several protocol variants have been proposed to improve practicality~\cite{2018Ma,2018Wang,2019Cui,2019Curty}, such as sending-or-not-sending (SNS)~\cite{2018Wang} and no-phase-post selection (NPP)~\cite{2019Cui,2019Curty}, which partially eliminate the need for phase-slice post-selection. Significant experimental advances have confirmed TF-QKD's superior rate-loss scaling~\cite{2019Minder,2020Chen,2021Pittaluga,2022Wang-830,2023liu}, with recent demonstrations extending beyond 1000 km~\cite{2023liu}. 

Similar to measurement-device-independent QKD~\cite{2012Lo}, participants in TF-QKD send quantum signals to an intermediate measurement node, enabling immunity to all potential side-channel attacks on measurement devices. This configuration is well-suited for star-type network expansion. However, the stringent requirements of twin-field phase tracking, typically involving additional service fibers and optical frequency-locking hardware, present challenges to building the scalability of TF-QKD networks. Recently, TF-QKD networks based on various TF-QKD network architechture including a simple ring architechture~\cite{2022Zhong} and a $2 \times N$ plug-and-play architechture ~\cite{2022Park} have been proposed to bypass using phase and frequency locking techonolgy. However, these network architectures require additional components to prevent security vulnerabilities associated with bidirectional optical paths~\cite{2006Gisin,2008Zhao,2010Zhao,2014Nitin}. Furthermore, to enable multi-user key distribution, time-division multiplexing of a single detection unit is required. This presents a significant challenge to the detector's bandwidth as the number of users increases, thereby reducing the achievable secure key rate per user.

In this study, we propose a scalable TF-QKD network with adaptable architecture. Our architecture is inspired by recent advancements in protocols~\cite{2022Zeng,2022Xie,2023Liu-advantage} and open-architecture TF-QKD schemes~\cite{2023Zhou-exp,2023Zhu,2023Li-TF,2024Zhou,2024Chen}, eliminating the need for additional channels and devices to synchronize the frequencies of independent light sources, offering greater flexibility and scalability. Furthermore, each user prepares quantum signals and sends them to a network node in single-path transmission, which inherently provides immunity to bidirectional transmission security vulnerabilities. Furthermore, the network node employs an optical switch to dynamically group users and, in conjunction with multi-user measurement units (MUs), flexibly supports secure key distribution for the network. Within multi-user MU, the output port of each user is expanded using a multi-port splitter, enabling simultaneous key generation among a larger number of user pairs.

We conduct a proof-of-principle demonstration of a three-user TF-QKD network, in which secure key exchange is successfully achieved between any pair of users over simulated link losses of up to $30$ dB, with an average rate of $5.01 \times 10^{-7}$ bit/pulse or $19.57$ bit/s. Additionally, we perform simulations based on the parameters of the experimental system. The results show that the proposed architecture can achieve a total secure key rate of $1.24 \times 10^{-3}$ bit/pulse or $4.84 \times 10^{4}$ bit/s at $100$ km in a symmetric $32$-user network with a specific MU configuration. This network architecture enhances the topology of untrusted-node QKD networks and promotes their practical applications.

    	\section{Network architecture}

        
        Figure~\ref{fig_network_architecture}(a) depicts the adaptive architecture of the TF-QKD network. The network adopts a star topology centered on the network node, with users located at the terminals of quantum channels extending from the node. Each user connects to the TF-QKD network as a transmitter that prepares quantum state according to specific TF-QKD protocols. The network node contains an $N \times N$ optical switch and multiple MUs. The optical switch groups arbitrary users and routes their quantum signals to the same MU, such as the $2$-user and $3$-user MUs. Figure~\ref{fig_network_architecture}(b) and (c) show $n$-user MU configurations featuring varying numbers of extended ports per user output. The MU expands each user’s output to support their connections with multiple users. The specific number of extended ports depends on user's demands. 
        
        \begin{figure*}[htp]
        	\centering
        	\includegraphics[width=0.90\linewidth]{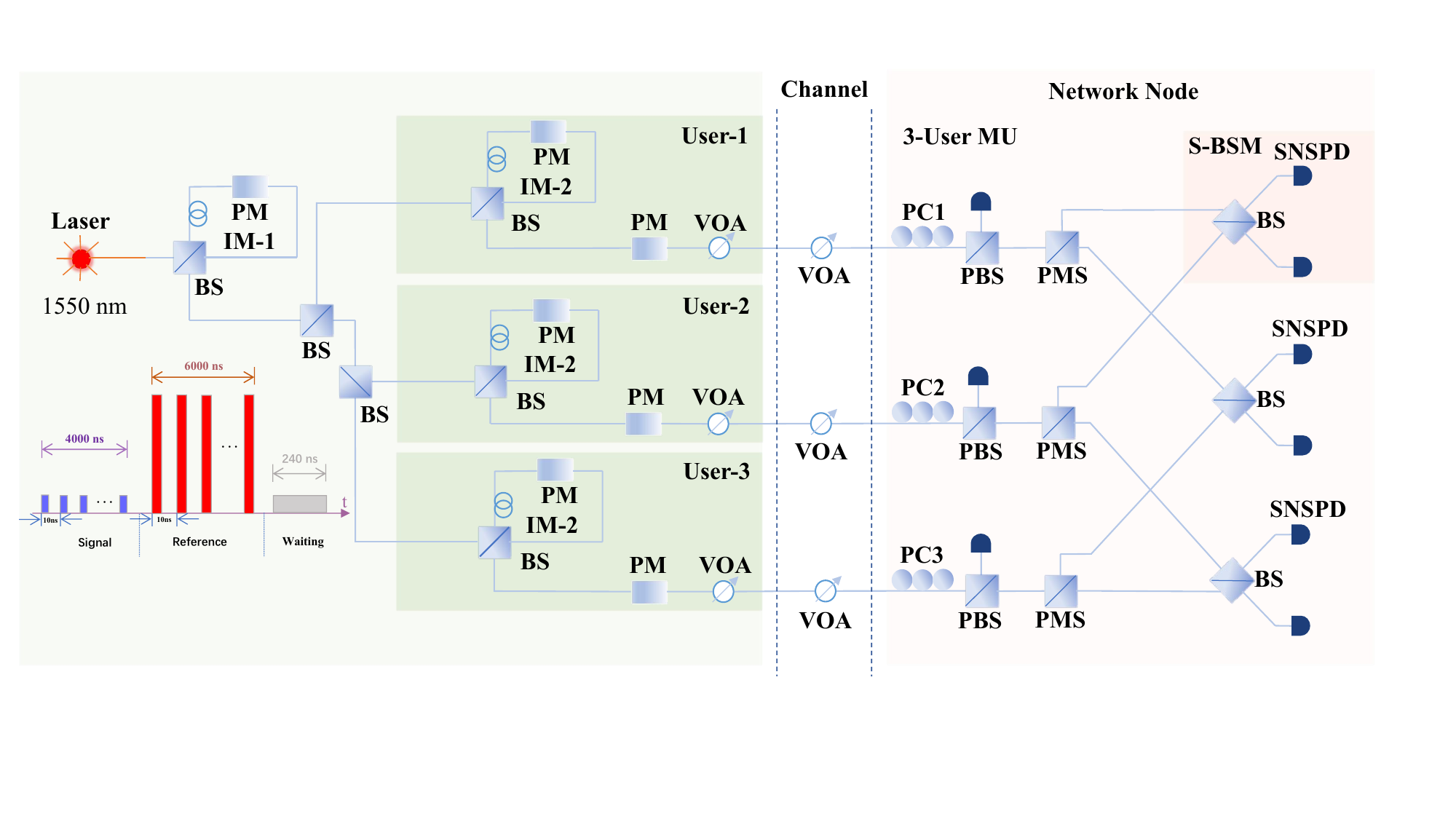}
        	\caption{Schematic of our experimental setup. A commercial laser serves as the light source for all users. Each frame of light pulse is modulated by the first intensity modulation module (IM-$1$) into $400$ signal pulses, $600$ reference pulses, and $24$ vacuum pulses. The intensity modulation module includes beam splitter (BS) and a phase modulator (PM). Subsequently, each user employs a second intensity modulation module (IM-$2$) and a phase modulator (PM) to implement encoding, decoy intensity modulation, and discrete random phase modulation on the optical pulses. The pulses are then attenuated via a variable optical attenuator (VOA), followed by a simulated channel using VOA. At the network node, quantum signals from user pass through polarization controller (PC) and polarization beam splitter (PBS), then couple into a polarization-maintaining splitter (PMS). The signals from the user pairs interfere at a BS within the single-photon Bell state measurement (S-BSM) module. Finally, the quantum signals are detected by superconducting nanowire single-photon detectors (SNSPDs).}
        	\label{fig_setup}
        \end{figure*}
        
        In this network, key distribution between any pair of users is implemented based on the TF-QKD protocol, which requires coherent control of the light fields from remote users. The frequency difference between independent user lasers or phase shifts induced by channel fluctuations affect the coherence of these light fields. Building on recent advancements~\cite{2023Li-TF}, users can employ  commercial kilohertz-linewidth semiconductor lasers to successfully perform TF-QKD. This scheme reconstructs a global phase reference through post-processing, eliminating the need for additional service fibers for frequency dissemination and thereby significantly simplifying hardware requirements and network deployments. Accordingly, the transmitter typically includes a commercial laser, multiple intensity modulators, and phase modulators. Furthermore, there are mature solutions to effectively compensate for the phase shifts caused by fiber fluctuations over several hundred kilometers, either through real-time feedback~\cite{2021Pittaluga,2022Wang-830,2023Zhou-exp} or post-processing techniques~\cite{2023liu,2023Li-TF,2024Chen}. 
        
        The network node employs an optical switch and multiple MUs to support any-to-any user connectivity and adapt to complex, real-world network topologies. Specifically, the reconfigurable optical switch enables adaptable grouping of users through multiple MUs, subject to the constraint $\sum_{2}^{n} n M_{n} < N$, where $N$ denotes the number of optical switch output ports, $n$ represents the output ports occupied by an $n$-user MU, and $M_{n}$ indicates the quantity of $n$-user MUs deployed. 
        
        For instance, a $2$-user MU essentially corresponds to the single-photon Bell state measurement (S-BSM) module used in typically TF-QKD system. A $3$-user MU extends each user's output to two ports, enabling simultaneous pairwise connectivity among all users. For a $n$-user MU, as shown in Fig.~\ref{fig_network_architecture}(b), each user’s output is extended to two ports, allowing the formation of up to $n$ distinct user pairs within the unit. Under this modular MU configuration, the total number of supported concurrent user pairs becomes $M_{2}+\sum_{3}^{n} n M_{n,2}$, where $M_{n,2}$ denotes the number of $n$-user MUs with two extended ports per user. Intuitively, increasing the number of output ports per user enables more user pairs to be supported simultaneously, enhancing adaptability to diverse connection demands in real-world scenarios. As shown in Fig.~\ref{fig_network_architecture}(c), an $n$-user MU with three output ports per user can support up to $\lfloor 3n/2 \rfloor$ pairs of users. Incorporating such MUs into the network increases the maximum number of concurrent connections to $M_{2} + \sum_{3}^{n} n M_{n,2} + \sum_{4}^{n} \lfloor 3n/2 \rfloor M_{n,3} $, where $M_{n,3}$ is the number of $n$-user MUs with three extended ports per user. 
        
        More generally, when MUs with varying numbers of extended ports are integrated, the network’s maximum concurrent connections $C$ can expressed as:
        \begin{equation}
        \begin{aligned}
         C = M_{2} + \sum_{3}^{n} n M_{n,2} + \sum_{4}^{n} \sum_{i=3}^{n-1} \lfloor ni/2 \rfloor M_{n,i},
        	\end{aligned}
        \end{equation}
        subject to the resource constraint $2M_{2} + \sum_{3}^{n} n \sum_{i=2}^{n-1} M_{n,i} < N $.
        Here, $M_{n,i}$ denotes the number of $n$-user MUs with $i$ extended ports per user. These modular MU configurations, when integrated with a reconfigurable optical switch, significantly enhance the network’s scalability and flexibility. They enable the system to dynamically allocate resources and satisfy varying user connection demands while adhering to the optical switch's total output port constraints. \blk 
	
     	\section{Experiment}\label{Experiment}
     	
     	\begin{figure*}[htp]
     		\centering
     		\includegraphics[width=0.65\linewidth]{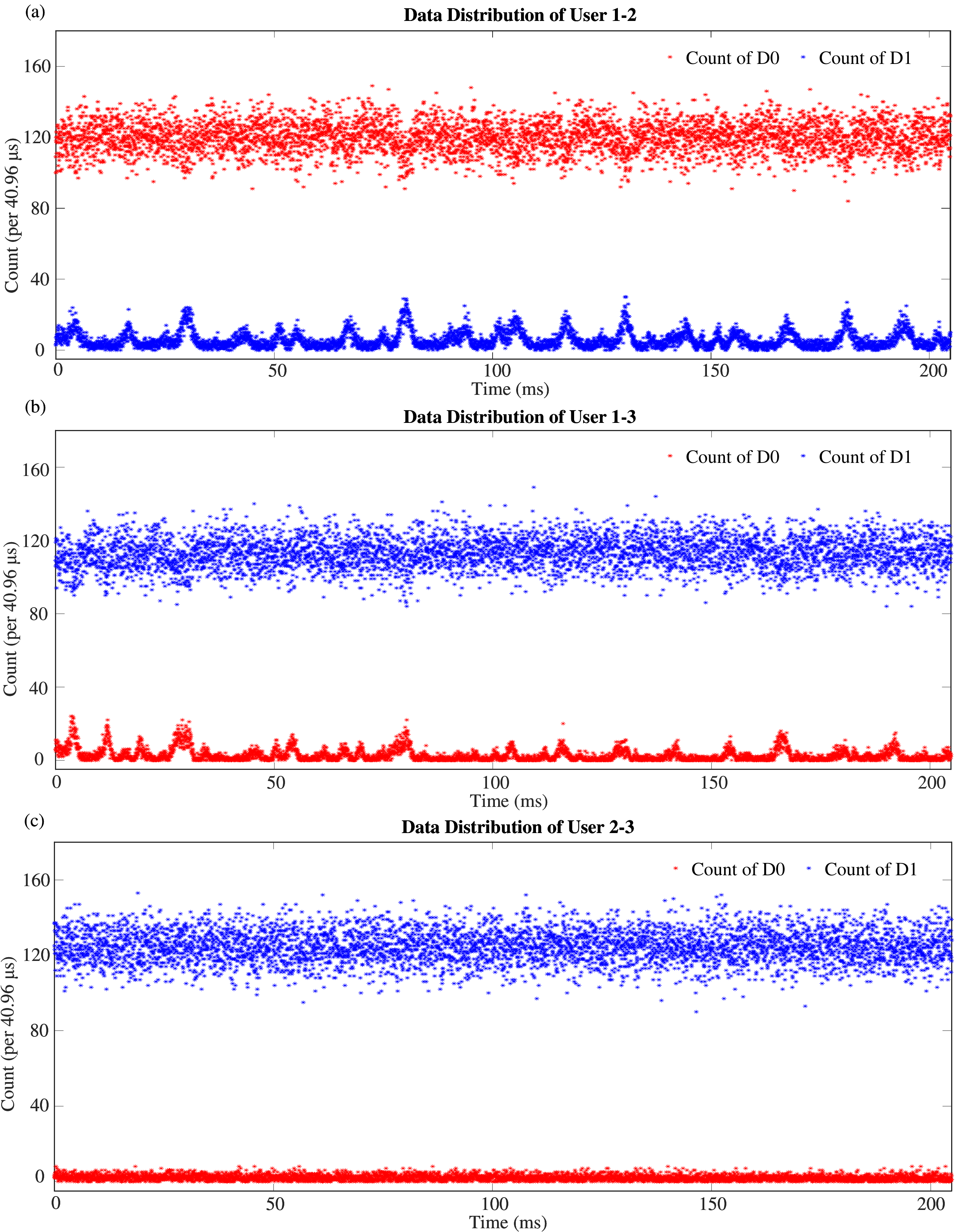}
     		\caption{Interference results between users in a three-user network within $200$ ms. The interference results are captured by recording the counts of the detectors D0 and D1. The average interference visibilities for user pairs $1$-$2$, $1$-$3$, and $2$-$3$ are $90.87\%$, $95.82\%$, and $96.72\%$, respectively.}
     		\label{fig_inter_result}
       \end{figure*} 
     
       We build a proof-of-principle experimental setup in which three users are statically connected to a three-user MU, employing a single-laser scheme rather than fully independent lasers. Despite this simplification, our experiment can serve as a functional instantiation of the adaptive network architecture components. Figure~\ref{fig_setup} shows the schematic of experimental setup. A laser source operating at $1550$ nm generates optical pulses at a $100$ MHz clock rate, with a pulse width of approximately $500$ ps. These pulses enter the first intensity modulation module (IM-$1$), which is a Sagnac-based interferometer. The optical pulses are modulated into sequential frame-based signals. Specifically, each frame consists of $1024$ optical pulses, comprising $400$ signal pulses, $600$ reference pulses and $24$ vacuum pulses. These pulses are then sent to the three users.
       
       Each user encodes the signal pulses according to the three-intensity sending-or-not-sending (SNS) TF QKD protocol~\cite{2018Wang,2019Yu,2021Jiang} by employing the second intensity modulation module (IM-$2$), thereby producing a signal window for key extraction and a decoy window to estimate information leakage. Subsequently, a phase modulator (PM) is employed to perform $16$-level random phase modulation over a range of $2\pi$ to the signal pulses. Finally, the signal pulses are attenuated to the single-photon level using a variable optical attenuator (VOA) before entering the channel. 
       
       \begin{figure*}[htp]
       	\centering
       	\includegraphics[width=0.85\linewidth]{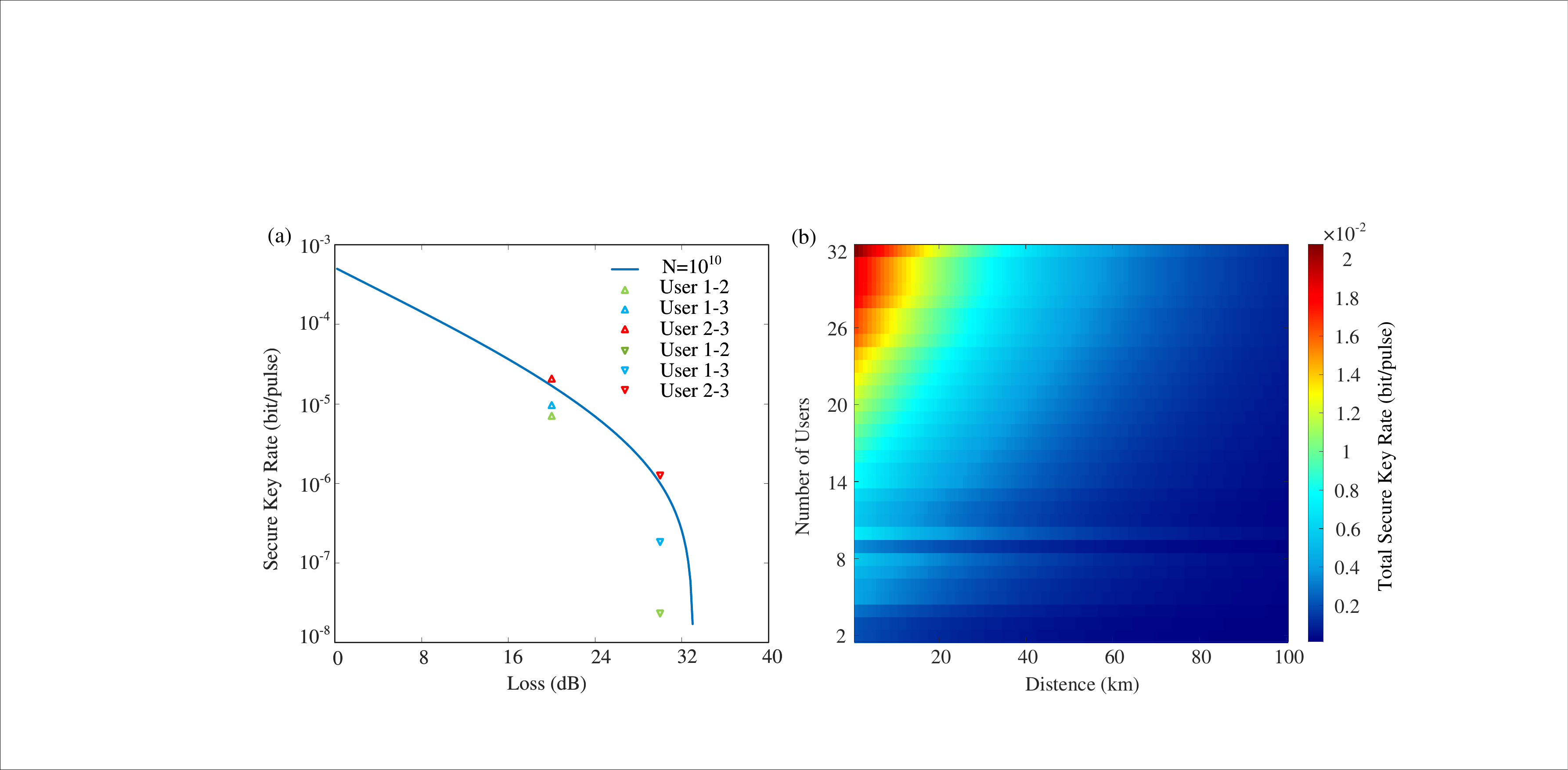}
       	\caption{(a) Experimental results in a three-user network. (b) The total secure key rate of the network over varying transmission distances and different numbers of active users.}
       	\label{fig_result_simulation}
       \end{figure*}  
       
       Signals from different users reach the three-user MU of network node through a channel simulated by a VOA. The arriving quantum signals are first processed by a polarization controller (PC) to align their polarization to the one of orthogonal modes of the followed polarization beam splitter (PBS). The signals are then efficiently coupled into a $1 \times 2$  fast-axis block polarization-maintaining splitter (PMS), and randomly routed to different branches with the same polarization. The signals from the user pairs interfere with the same polarization at the beam splitter (BS) within the S-BSM module. The interference outputs between user pairs are detected by two superconducting nanowire single-photon detectors (SNSPDs), with detection efficiency around $45\%$ and a dark count rate of about $8\times10^{-8}$. The detected photon events are recorded by the TDC and then processed by a computer. The detection events of the reference pulse are used in post-processing to estimate the phase fluctuation between users. Within the effective arrival time $t$, the detection events in the decoy window are further filtered according to the following condition:     
     
       \begin{equation}
       \begin{aligned}
     	\left|\cos \left(\theta_{i}-\theta_{j}-\varphi_{i j}\right)\right| \geq \cos (\pi / 16),
     	\end{aligned}
     	\label{phase_condition}
       \end{equation}	
       where $\theta_{i}$ ($\theta_{i}$) is the encoded phase of user-$i$ ($j$) to be announced in post-processing, and $\varphi_{i j}$ is the phase fluctuation between users. We also apply the actively odd-parity pairing (AOPP) method~\cite{2021Jiang} to suppress bit-flip errors and improve the secure key rate. 
    	
    	\section{Results}
    	
    	Using the described experimental system, we first measure the interference between users to preliminarily assess system performance. The measurement is conducted with the phase modulator installed but not actively modulating. The interference results are obtained by monitoring the counts from detectors in the MU, as shown in Figure~\ref{fig_inter_result}. We observe that the average interference visibilities for user pairs $1$-$2$, $1$-$3$, and $2$-$3$ are $90.87\%$, $95.82\%$, and $96.72\%$, respectively, over a period of approximately $200$ ms. Due to the single-laser scheme adopted, phase fluctuations affecting interference visibility in the experiment primarily originate from optical components.
        
        We then perform a three-user TF-QKD network experiment. The transmittance for the users is simulated using optical attenuators, which model the transmittance of fiber links with lengths of $100$ km and $150$ km, assuming a loss coefficient of $0.2$ dB/km. Based on the similar transmittance of all users, we select the same implementation parameters for simplicity. Considering the finite-key effect, the secure key rate of after AOPP is calculated using ~\cite{2021Jiang}       
          
        \begin{equation}
        	\begin{aligned}
             R = & \frac{1}{N} \{ {n_{1}^{\prime} (1-h(e_{1}^{\prime ph}))-f n_{t}^{\prime} h(E^{\prime})} 
             \\ & {-2 \log_{2} \frac{2}{\varepsilon_{cor}}-4 \log_{2} \frac{1}{\sqrt{2}\varepsilon_{PA} \hat{\varepsilon}}} \},
        	\end{aligned}
        	\label{l_finite}
        \end{equation}
        where $N$ is the total number of send pulse pairs, $h(x)= -x \log_{2} x-(1-x) \log_{2}(1-x)$ is the Shannon entropy function. $f$ is the error correction efficiency equal to $1.1$. $n_{t}^{\prime}$ is the number of remaining bits after AOPP. $E^{\prime}$ is the bit-flip error rate of the remaining bits after AOPP. $\varepsilon_{cor}$ is the failure probability of error correction, $\varepsilon_{PA}$ is the failure probability of privacy amplification, $\hat{\varepsilon}$ is the coefficient obtained using the chain rules of smooth min- and max-entropy, with all three parameters being set to $10^{-10}$. Detailed parameter estimation for the three-intensity SNS-TF QKD with AOPP protocol can be found in Appendix~\ref{Appendix_A}. 
        
        For each transmittance level, a total of $10^{10}$ signal pulses are emitted by each user. The detection events of the reference pulses are extracted within a time window of $200$ ms, and only the data with stable interference are retained. We estimate phase fluctuation from the reference data during this time interval and apply these values to compensate for the phase fluctuation of signal pulses for secure key generation. As shown in Figure~\ref{fig_result_simulation}(a), the secure key rate for user pairs $1$-$2$, $1$-$3$, and $2$-$3$ is respectively $2.38 \times 10^{-8}$ ($6.87 \times 10^{-6}$), $1.88 \times 10^{-7}$ ($9.35 \times 10^{-6}$), and $1.29 \times 10^{-6}$ ($2.02 \times 10^{-5}$) bits/pulse under a simulated channel loss of $30$ ($20$) dB. Detailed implementation parameters and measurement results are summarized in Appendix~\ref{Appendix_B}. 
        
        To evaluate the network’s performance with a larger number of users, we conduct numerical simulation of a network that can support up to $32$ users in a symmetric scenario. We consider our system parameters along with the loss value of $1 \times 8$ splitter from reference~\cite{2013Frohlich}. The simulation assumes a specific MU configuration of $\{M_{2}=1, M_{3,2}=1, M_{4,2}=1, M_{6,2}=1, M_{8,2}=1, M_{9,8}=1\}$ and a total of $N=10^{11}$ pulses sent by each user. The total secure key rate of the network is evaluated with respect to transmission distance and the number of active users, as depicted in Fig.~\ref{fig_result_simulation}(b). As the number of active users increases, we determine the maximum number of  user pairs that can be simultaneously supported under the given MU configuration, and calculate the corresponding total secure key rate of the network. 
        
        The independent S-BSMs within the MUs at the network node improve the overall network’s capacity for parallel key generation. The fully loaded network achieves a total secure key rate of $1.24 \times 10^{-3}$ bit/pulse or $4.84 \times 10^{4}$ bit/s at a transmission distance of $100$ km. The achievable secure key rate between user pairs can reach up to $4.77 \times 10^{3}$ bit/s, which meets the bit rate requirement for real-time one-time-pad encryption of voice communications. Even for the worst-performing user pairs, a secure key rate of approximately $256.86$ bit/s can be achieved under the quantum transmission duty cycle used in our experiment.
        
        With nine active users, the $9$-user MU with eight extended ports enables up to $28$ pairs of users simultaneously. Although this MU allows for a greater number of concurrent connections, it also requires significantly more detection resources. Moreover, due to the higher loss in multi-port MUs, the key rate of each user pair becomes more susceptible to statistical fluctuations at the same transmission distance, thereby limiting the network performance, as shown in Fig.~\ref{fig_result_simulation}(b). Therefore, it is crucial to optimize the MU composition based on practical resource constraints and user demands to achieve a balance between flexibility, resource usage, and network performance.

     	\section{Conclusion} 
        
        In summary, we propose a scalable TF-QKD network that supports adaptive user pairing based on a modular design. By integrating an $N \times N$ optical switch with multi-user MUs of varying port configurations, our network supports flexible connectivity and concurrent key generation among multiple users. We perform a proof-of-principle network experiment with a three-user MU and demonstrate secure key distribution between any pairs of users over simulated fiber losses up to $30$ dB. Furthermore, numerical simulation shows that the proposed architecture can achieve a total secure key rate of $4.84 \times 10^{4}$ bit/s over a distance of $100$ km in a $32$-user network.  
        
        This work advances the practical deployment of untrusted-node QKD networks, offering a scalable and efficient solution for future quantum-secured communication infrastructures. Furthermore, this architecture, combined with compact integrated chip-based systems~\cite{2020Wei} and cost-effective hybrid quantum-classical system~\cite{2024pittaluga}, is expected to further drive the practical application of QKD network. Beyond QKD, this architecture could be adapted into other quantum protocols, such as quantum digital signature~\cite{2025Du} and quantum blockchain~\cite{2023Weng}.  
      
       \begin{table*}[!htbp]
      	\centering
      	\caption{The three-intensity parameters under different simulated losses, including intensities and probabilities. $\{\mu_{o}, \mu_{x}, \mu_{y}\}$ represent three sources with different intensities. $p_x$ is the probability of sending decoy intensity $\mu_{x}$. $p_y$ is the probability of sending the signal window. $\varepsilon$ is the probability of sending during the signal window. $\mu_{ref}$ is the intensity of reference pulse.}
      	\doublerulesep 2pt \tabcolsep 10pt
      	\renewcommand\arraystretch{1.3}
      	\begin{tabular}{cccccccc}
      		\hline
      		\hline
      		$Loss$ (dB) & $\mu_y$ & $\mu_x$ & $\mu_o$   & $p_x$   & $p_y$  & $\varepsilon$ & $\mu_{ref}$  \\  \hline
      		$20$      & $0.44$  & $0.01$  & $0.0016$  & $0.23$  & $0.72$ & $0.25$        & $1.5$        \\  \hline
      		$30$      & $0.43$  & $0.01$  & $0.0016$  & $0.36$  & $0.53$ & $0.25$        & $1.5$         \\ \hline  
      	\end{tabular}\label{operation_parameter}
      \end{table*}

	    \section*{Acknowledgment}
    	This study was supported by the National Natural Science Foundation of China (Nos. 62171144 and 62031024), Guangxi Science Foundation (Nos. 2025GXNSFAA069137 and GXR-1BGQ2424005), and Innovation Project of Guangxi Graduate Education (No. YCBZ2025064).

	    \appendix 
    	\section{Three-intensity SNS-TF QKD with active odd parity pairing (AOPP) protocol}\label{Appendix_A}
    	
    	Here, we describe the three-intensity SNS-TF QKD protocol~\cite{2018Wang,2019Yu} adopted in our network demonstration, where the AOPP method is employed to extract the final secret key. Given that user-$i$ and user-$j$ in the network request key distribution session. User-$i(j)$ possesses three sources with different intensities, denoted as $\mu_{o_i}, \mu_{x_i}, \mu_{y_i} (\mu_{o_j}, \mu_{x_j}, \mu_{y_j}) $, where $\mu_{o_i}(\mu_{o_j})$ represents vacuum source. In each time window, user-$i(j)$ randomly determines whether the optical pulse originates from the vacuum source $\mu_{o_i}(\mu_{o_j})$ with probability $p_o$, the decoy source $\mu_{x_i}(\mu_{x_j})$ with probability $p_x$, or a signal window with probability $p_y=1-p_o-p_x$. If a signal window is determined, user-$i(j)$ then randomly chooses between the source $\mu_{o_i}(\mu_{o_j})$ with probability $ 1-\epsilon $ and the source $\mu_{y_i}(\mu_{y_j})$ with probability $ \epsilon $. For vacuum source in the signal window, user-$i(j)$ encodes bit $0(1)$, whereas for source $\mu_{y_i}(\mu_{y_j})$ in the signal window, user-$i(j)$ encodes bit $1(0)$.
    	
    	Considering a key distribution session, the total number of pulses sent by user-$i(j)$ is $N$. The number of source pairings $lr (l \in \{\mu_{o_i}, \mu_{x_i}, \mu_{y_i}\}, r \in \{\mu_{o_j}, \mu_{x_j}, \mu_{y_j}\})$ can be represented as $N_{lr}$, and the total number of corresponding single detector response events is represented as $n_{lr}$. The yield of source $lr$
    	can be defined as $S_{lr} = n_{lr}/N_{lr}$, and the expected value is represented as $\left \langle S_{lr} \right \rangle $. Based on the above definitions, we have:
    	
    	\begin{equation}
    	\begin{array}{c}
    		N_{\mu_{o_{i}} \mu_{o_{j}}}=\left[p_{o}^{2}+2 p_{o} p_{y}(1-\varepsilon)\right] N, \\
    		N_{\mu_{o_{i}} \mu_{x_{j}}}=N_{\mu_{x_{i}} \mu_{o_{j}}}=\left[p_{o}+p_{y}(1-\varepsilon)\right] p_{x} N, \\
    		N_{\mu_{o_{i}} \mu_{y_{j}}}=N_{\mu_{y_{i}}} \mu_{o_{j}}=p_{o} p_{y} \varepsilon N .
    	\end{array}
    	\end{equation}
    	
    	For the time window in which user-$i$ and user-$j$ select source $\mu_{x_i}$ and $\mu_{x_j}$, the users will announce the phases encoded and perform post-selection on the single detector response events based on the following conditions:
    	
    	\begin{equation}
    	\begin{aligned}
    	 	1 - \left|\cos \left(\theta_{i}-\theta_{j}-\varphi_{i j}\right)\right| \le \lambda,
    	\end{aligned}
    	\end{equation}
    	where $\theta_{i}$ and $\theta_{j}$ are the encoding phase of the pulses for user-$i$ and user-$j$ respectively, $\varphi_{ij}$ the phase fluctuation between user-$i$ and user-$j$, and $\lambda$ is a small positive value. Let the number of source pairings satisfying the above condition be denoted as $N_{x}$, and the number of error events as $m_{x}$. Then, the error rate for the window $\mu_{x_i}\mu_{x_j}$ us given by $T_{x} = m_{x} / N_{x} $.
    	
    	To obtain the secure key rate, the users need to estimate the lower bound of the number of untagged bits and the upper bound of the phase error rate of the untagged bits. After performing decoy state analysis, the lower bound of the expected counting rate for the states $|01\rangle$$\langle01|$ and $|10\rangle$$\langle10|$ can be expressed as follows:
    	
    	\begin{footnotesize}
    	\begin{equation}
    	\begin{aligned}
    		\underline{\left\langle s_{01}\right\rangle} & = \frac{\mu_{y}^{2} e^{\mu_{x}} \underline{\left\langle S_{\mu_{o_{i}} \mu_{x_{j}}}\right\rangle}-\mu_{x}^{2} e^{\mu_{y}} \overline{\left\langle S_{\mu_{o_{i}} \mu_{y_{j}}}\right\rangle}-\left(\mu_{y}^{2}-\mu_{x}^{2}\right) \overline{\left\langle S_{\mu_{o_{i}} \mu_{o_{j}}}\right\rangle}}{\mu_{y} \mu_{x}\left(\mu_{y}-\mu_{x}\right)},\\   		
    		\underline{\left\langle s_{10}\right\rangle} & = \frac{\mu_{y}^{2} e^{\mu_{x}} \underline{\left\langle S_{\mu_{x_{i}} \mu_{o_{j}}}\right\rangle}-\mu_{x}^{2} e^{\mu_{y}} \overline{\left\langle S_{\mu_{y_{i}} \mu_{o_{j}}}\right\rangle}-\left(\mu_{y}^{2}-\mu_{x}^{2}\right) \overline{\left\langle S_{\mu_{o_{i}} \mu_{o_{j}}}\right\rangle}}{\mu_{y} \mu_{x}\left(\mu_{y}-\mu_{x}\right)},
    	\end{aligned}{\tiny} 
    	\end{equation}
        \end{footnotesize}
    	then the lower bound of the counting rate for the untagged bits can be expressed as follows:
    	
    	\begin{equation}
    	\begin{aligned}
    		\underline{\left\langle s_{1}\right\rangle} = \frac{1}{2}(\underline{\left\langle s_{01}\right\rangle} +    \underline{\left\langle s_{10}\right\rangle}).
    	\end{aligned}
    	\end{equation}
    	
    	The lower bound of the expected values for the number of untagged bit $1$ and untagged bit $0$ are given by: 
    	
    	\begin{equation}
    	\begin{aligned}
           \underline{\left\langle n_{10}\right\rangle} = N p_{y}^{2} \epsilon(1-\epsilon) \mu_{y} e^{-\mu_{y}} \underline{\left\langle s_{10}\right\rangle}, \\ 
           \underline{\left\langle n_{01}\right\rangle} = N p_{y}^{2} \epsilon(1-\epsilon) \mu_{y} e^{-\mu_{y}} \underline{\left\langle s_{01}\right\rangle}.
    	\end{aligned}
    	\end{equation}
    	
    	The upper bound of the expected values for the phase error rate $1$ is given by:
    	
    	\begin{equation}
    	\begin{aligned}
           \overline{\left\langle e_1^{ph} \right\rangle} = \frac{\left\langle T_x \right\rangle - e^{-2\mu_{x}} \left\langle S_{\mu_{o_{i}} \mu_{o_{j}}} \right\rangle / 2}{2\mu_{x} e^{-2\mu_{x}} \underline{\left\langle s_1 \right\rangle}}.
    	\end{aligned}
    	\end{equation}
    	
    	        \begin{table*}[!htbp]
    		\centering
    		\caption{Detailed experimental results under different simulated losses, including the secure key rate considering finite-key effect and AOPP. The detection events are labeled as "Detected-$ijab$", where '$i(j)$' and '$a(b)$' represent the signal $Z$ or decoy $X$ time windows and intensities determined by user-$i(j)$.}
    		\doublerulesep 0.2pt \tabcolsep 8pt
    		\renewcommand\arraystretch{1.3} 
    		\begin{tabular}{c|ccc|cccc}
    			\hline
    			\hline
    			{}                           & User$1$-$2$                  & User$1$-$3$             & User$2$-$3$             & User$1$-$2$            & User$1$-$3$               & User$2$-$3$                  &  \\ \hline
    			Loss (dB)                    & \multicolumn{3}{c|}{20}                                                          & \multicolumn{3}{c}{30}                                                             &  \\ \hline
    			$N$                          & \multicolumn{3}{c|}{$10^{10}$}                                                   & \multicolumn{3}{c}{$10^{10}$}                                                      &  \\ \hline
    			$R$ (bits/pulse)             & $6.8 \times 10^{-6}$         & $9.35 \times 10^{-6}$   & $2.02 \times 10^{-5}$   & $2.38 \times 10{-8}$   & $1.88 \times 10^{-7}$  & $1.29 \times 10^{-6}$              &  \\ \hline
    			Detected $ZZyy$              & 3984035                      & 4077966                 & 3645054                 & 854517                 & 601800                    & 648818                       &  \\ 
    			Detected $ZZoy$              & 6088650                      & 5973265                 & 5217181                 & 1022963                & 954205                    & 912414                       &  \\ 
    			Detected $ZZyo$              & 5160981                      & 4848325                 & 6250619                 & 912024                 & 751293                    & 1024881                      &  \\ 
    			Detected $ZZoo$              & 147816                       & 85275                   & 105192                  & 12931                  & 11578                     & 16709                        &  \\ 
    			Detected $ZXyx$              & 2348664                      & 2261991                 & 2816486                 & 791683                 & 1046887                   & 1003973                      &  \\ 
    			Detected $ZXox$              & 166084                       & 182167                  & 213010                  & 95600                  & 32038                     & 71703                        &  \\ 
    			Detected $ZXyo$              & 445273                       & 594330                  & 552265                  & 179775                 & 359825                    & 206623                       &  \\ 
    			Detected $ZXoo$              & 14484                        & 6656                    & 12434                   & 5705                   & 3344                      & 6423                         &  \\ 
    			Detected $XZxy$              & 2421535                      & 1853833                 & 2490897                 & 992859                 & 983220                    & 784572                       &  \\ 
    			Detected $XZoy$              & 587264                       & 501135                  & 407128                  & 349189                 & 202943                    & 228031                       &  \\ 
    			Detected $XZxo$              & 211380                       & 192685                  & 140001                  & 34267                  & 97368                     & 70640                        &  \\ 
    			Detected $XZoo$              & 15243                        & 7888                    & 12356                   & 7475                   & 2976                      & 4317                         &  \\ 
    			Detected $XXxx$              & 130416                       & 180341                  & 127632                  & 103282                 & 104274                    & 145071                       &  \\ 
    			Detected $XXox$              & 16158                        & 13966                   & 24151                   & 23824                  & 12949                     & 22445                        &  \\
    			Detected $XXxo$              & 18007                        & 14983                   & 12346                   & 13302                  & 25180                     & 21872                        &  \\ 
    			Detected $XXoo$              & 350                          & 788                     & 475                     & 702                    & 1518                      & 651                          &  \\ \hline
    			Detected $XXxx$-accepted     & 19109                        & 26176                   & 20232                   & 17799                  & 19131                     & 23062                        &  \\ 
    			Correct $XXxx$-accepted      & 17381                        & 24176                   & 18694                   & 16538                  & 17859                     & 21208                        &  \\ 
    			QBER $ZZ$ before AOPP        & 26.86\%                      & 27.78\%                 & 24.64\%                 & 30.95\%                & 26.45\%                   & 25.57\%                      &  \\ 
    			QBER $ZZ$ after AOPP         & 1.84\%                       & 1.19\%                  & 1.16\%                  & 1.16\%                 & 0.95\%                    & 1.13\%                       & 
    			\\ \hline
    		\end{tabular}\label{experimental_result}
    	\end{table*}
    	
    	We further calculate the lower bounds on the number of untagged bit $ n_1' $ and the phase error rate $ e_1'^{ph} $ after AOPP according to the methods in reference~\cite{2021Jiang}. The related formulas of $n_1'$ are as follows:
    	
    	\begin{equation}
    	\begin{aligned}
    		u & = \frac{n_g}{2n_{odd}}, \\
    		\underline{n_{10}} & = \varphi^L(\underline{\langle n_{10} \rangle}), \\
    		\underline{n_{01}} & = \varphi^L(\underline{\langle n_{01} \rangle}), \\
    		\underline{n_1} & = \underline{n_{01}} + \underline{n_{01}}, \\
    		n_1^r & = \varphi^L\left( \frac{\underline{n_1}}{n_t} \frac{\underline{n_1}}{n_t} \frac{u n_t}{2} \right), \\
    		n_{10}' & = 2 n_1^r \left( \frac{\underline{n_{10}}}{\underline{n_1}} - \sqrt{\frac{-\ln \epsilon}{2 n_1^r}} \right), \\
    		n_{01}' & = 2 n_1^r \left( \frac{\underline{n_{01}}}{\underline{n_1}} - \sqrt{\frac{-\ln \epsilon}{2 n_1^r}} \right), \\
    		n_{min} & = \min(n_{01}', n_{10}'), \\
    		n_1' & = 2 \varphi^L \left( n_{min} \left( 1 - \frac{n_{min}}{2 n_1^r} \right) \right),
    	\end{aligned}
    	\end{equation}
    	where $n_t$ is the raw key generated by users-$i$ and user-$j$, $n_g$ is  the number of pair if users-$i$ and user-$j$ perform AOPP to their raw keys. $n_{odd}$ is the number of pair with odd-parity when user-$j$ randomly groups their raw keys two by two. $\epsilon$ is the failure probability of parameter estimation. $\varphi^U(x)$ and $\varphi^L(x)$ are the upper and lower bounds after using Chernoff bound~\cite{1952Chernoff} to estimate the real values based on the expected values.
    	
    	The related formulas of $ e_1'^{ph} $ are as follows:
    	
    	\begin{equation}
        \begin{aligned}
    		r &= \frac{\underline{n_1}}{\underline{n_1} - 2n_1^r} \ln \frac{3(\underline{n_1} - 2n_1^r)^2}{\epsilon}, \\
    		e_{\tau} &= \frac{\varphi^{U}(2n_1^r \overline{\langle e_1^{ph}\rangle})}{2n_1^r - r}, \\
    		\overline{M_s} &= \varphi^{U}((n_1^r - r) e_{\tau} (1 - e_{\tau})) + r, \\
    		e_1'^{ph} &= \frac{2\overline{M_s}}{n_1'}.
        \end{aligned}
    	\end{equation}
    	
    	Finally, the secure key rate can expressed by:
    	       
    	\begin{equation}
        \begin{aligned}
    			R = & \frac{1}{N} \{ {n_{1}^{\prime} (1-h(e_{1}^{\prime ph}))-f n_{t}^{\prime} h(E^{\prime})} 
    			\\ & {-2 \log_{2} \frac{2}{\varepsilon_{cor}}-4 \log_{2} \frac{1}{\sqrt{2}\varepsilon_{PA} \hat{\varepsilon}}} \}.
        \end{aligned}
    	\end{equation}
    	 
    	\section{Detailed experimental results}\label{Appendix_B}
    	
    	Here, we summarize the parameters of the three-intensity SNS protocol used in our experiment, as shown in Table~\ref{operation_parameter}. Table~\ref{experimental_result} presents detailed experimental results under different simulated losses, including the single-detector response events used to calculate the secure key rate and the AOPP results.

        \end{document}